\documentclass[12pt]{article}
\usepackage{makeidx}
\usepackage{amssymb}
\usepackage{amsfonts}
\usepackage{amsmath}
\usepackage{geometry}
\usepackage[colorlinks=true, linkcolor=red, citecolor=blue]{hyperref}
\usepackage{float}
\usepackage{mathtools}
\usepackage{caption}
\usepackage{graphicx}
\usepackage[font=small,labelfont=bf,labelsep=space]{caption}
\usepackage{subfig}
\usepackage{setspace}

\makeatletter
\renewcommand\theequation{\thesection.\arabic{equation}}
\@addtoreset{equation}{section}
\makeatother

\input{tcilatex}
\begin{document}

\title{Zernike polynomials from the tridiagonalization of the radial
harmonic oscillator in displaced Fock states}
\author{Hashim A. Yamani${}^{\ast }$ and Zouha\"{\i}r Mouayn$^{\diamond
,\natural }$ \\
$^{\ast }${\footnotesize \ Dar Al-Jewar, Knowledge Economic City, Medina,
Saudi Arabia\vspace*{-0.2em}}\\
{\footnotesize \ \ e-mail: hashim.haydara@gmail.com \vspace*{0.6mm}}\\
{\footnotesize \ }$^{\diamond }${\footnotesize \ Faculty of Sciences and
Technics (M'Ghila),\vspace{-0.2em} P.O. Box. 523, B\'{e}ni Mellal, Morocco.}%
\\
{\footnotesize \ }$^{\natural }${\footnotesize Institut des Hautes \'{E}%
tudes Scientifiques, Paris-Saclay University,}\\
{\footnotesize \ 35 route de Chartres, 91893 Bures-sur-Yvette, France.}\\
{\footnotesize \ e-mail: mouayn@gmail.com, mouayn@ihes.fr}}
\maketitle

\begin{abstract}
We revisit the $J$-matrix method for the one dimensional radial harmonic
oscillator (RHO) and construct its tridiagonal matrix representation within
an orthonormal basis $\phi _{n}^{\left( z\right) }$ of $L^{2}\left( \mathbb{R%
}_{+}\right) ,$ parametrized by a fixed $z$ in\ the complex unit disc $%
\mathbb{D}$ and $n=0,1,2,...$ . Remarkably, for fixed $n$, \ and varying $%
z\in \mathbb{D}$, the system $\phi _{n}^{\left( z\right) }$ forms a family
of Perelomov-type coherent states associated with the RHO. For each fixed $n$%
, the expansion of $\phi _{n}^{\left( z\right) }$ over the basis $\left(
f_{s}\right) $ of eigenfunctions\ of the RHO yields coefficients $c_{n,s}(z,%
\overline{z})$ precisely given by two-dimensional complex Zernike
polynomials. The key insight is that the algebraic tridiagonal structure of
RHO contains the complete information about the bound state solutions of the
two-dimensional Schr\"{o}dinger operator describing a charged particle in a
magnetic field (of strength proportional to $B>%
{\frac12}%
)$ on the Poincar\'{e} disc $\mathbb{D}$.
\end{abstract}

\section{Introduction}

Zernike polynomials were originally introduced by F. Zernike \cite{Zer}, as
orthogonal polynomials defined over the unit disc and depending on two
parameters $\left( \alpha ,\beta \right) $ specifically for $\alpha =\beta
=1/2$. Later, in collaboration with Brinkman \cite{ZB}, they were extended
them to a more general form. They have since been generalized in many ways
and have found numerous applications in both mathematics and physics and
recently attracted more interested for their potential application in
biology.

\medskip

Actually, these polynomials were employed \cite{Fol} to expand the
Poisson-Szeg\H{o} kernel for the unit ball in $\mathbb{C}^{d}$. A related
Banach algebra was studied in \cite{Kan}. Additionally, they serve as
spherical functions for the homogeneous space $U\left( d\right) /U\left(
d-1\right) $, $\ d=2,3,\ldots $, see \cite{Ike},\cite{Koor1}, \cite{VK}, 
\cite{Wun}. In \cite{TNH} the authors have explored the coherent state
quantization of the unit disk using $2D$ Zernike polynomials$.$ They also
discussed the associated reproducing kernels and Hilbert spaces associated
with these polynomials. Their extension with respect to an extra parameter
was introduced in \cite{MZ}. \ More recently, the authors \cite{AS} while
formulating Donoho-Logan sieve principles for the wavelet transform on the
Hardy space on the Poincar\'{e} upper half-plane, explicit calculations
revealed a connection with the Zernike polynomials on the disc.

\medskip

In physics, \cite{BW} provides a comprehensive discussion on Zernike
polynomials, their optical applications. \ In this respect, \cite{Mah}
explores their complex representation in the context of optical aberrations
of systems with circular pupils. \cite{WC} explains wavefront aberrations
using the complex formalism of \ these polynomials., see also \cite{Wol}.
Beyond optics, they have been applied in structural biology. For instance, 
\cite{MRT} demonstrates their utility in identifying protein-protein binding
regions through their 2D polynomial expansion in the complex plane. We also
refer to the survey by \cite{NT} which provides an extensive bibliography on
these polynomials.

\medskip

On an other hand, tridiagonalization techniques, as explored in \cite{YM1},
particularly in the context of supersymmetry suggests the tridiagonalization
of the system Hamiltonian in displaced Fock states, where the first state
corresponds to a coherent state labeled by the complex coherence parameter $%
z $ belonging to the phase space$.$ These states, along with the system's
energy eigenstates, can be expressed in terms of one another. Interestingly,
the expansion coefficients are two-parameter complex functions with
intriguing properties. Indeed, in \cite{YM2} we have been dealing with the $%
J $-matrix method \cite{YF}, \cite{AYHA} for the harmonic oscillator
Hamiltonian, say $H_{har},$ to write down its tridiagonal matrix
representation in an orthonormal basis of $L^{2}\left( \mathbb{R}\right) $.
We rederived a set of generalized coherent states (GCS) of Perelomov type
labeled by points $z$ of the phase space $\mathbb{C}$ and depending on an
integer number $n\in \mathbb{Z}_{+}$, obtained as orbits of the operators of
the Schr\"{o}dinger representation of the Heisenberg group on a $n$th
Hermite function. Beside, the number states expansion in the basis of
eigenstates of $H_{har}$ gives rise to coefficients that are complex Hermite
polynomials $H_{n,s}\left( z,\overline{z}\right) $ \cite{Ism} whose linear
superpositions provide eigenfunctions for the two-dimensional magnetic
Laplacian associated with the $n$th Euclidean Landau level.

\medskip

In the present paper, we explore analogous questions for the radial harmonic
oscillator (RHO). Specifically, we revisit the $J$-matrix method for the RHO
and construct its tridiagonal matrix representation within an orthonormal
basis $\phi _{n}^{\left( z\right) }$ of $L^{2}\left( \mathbb{R}_{+}\right) $%
, where $z$ is a fixed parameter in\ the complex unit disc $\mathbb{D}$ and $%
n=0,1,2,...$ . We then demonstrate that for each fixed $n$, \ and varying $%
z\in \mathbb{D}$, the expansion of the state $\phi _{n}^{\left( z\right) }$
over the eigenstates of the RHO yields coefficients\ $c_{m,s}(z,\overline{z})
$\ that correspond exactly to two-dimensional complex Zernike polynomials.\
This expansion may also reflect the coherent state nature of $\phi
_{n}^{\left( z\right) }$ within the Hilbertian probabilistic framework \cite%
{Gaz}$.$ To further clarify this structure, we derive a closed-form
expression for the wavefunction of $\phi _{n}^{\left( z\right) }$. This
allows us to verify that the system $\left\{ \phi _{n}^{\left( z\right)
},z\in \mathbb{D}\right\} $ indeed forms a set of Perelomov-type coherent
states \cite{Per} obtained as orbits of operators from a square-integrable
representation of the affine group on a $n$th Laguerre function, or
equivalently by displacing a Fock state.

\medskip

A key insight of our work is that the algebraic tridiagonal structure of the
one-dimensional RHO operator encodes complete information about the bound
state solutions of the two-dimensional Schr\"{o}dinger operator describing a
charged particle in a magnetic field on the Poincar\'{e} disc. More
precisely, for $m\in \mathbb{Z}_{+}\cap \left[ 0,B-%
{\frac12}%
\right] $, linear superpositions the obtained\textit{\ }coefficients $%
c_{m,s}(z,\overline{z})$ generate eigenstates corresponding to the $m$th
hyperbolic Landau level, with magnetic field strength proportional to $B>%
{\frac12}%
.$ This approach provides an alternative solution in hyperbolic geometry to
the problem posed by the authors of \cite{GL} in the Euclidean setting,
where their analysis primarily relied on the cross-Wigner transform to
unitarily intertwine the Landau Hamiltonian with the harmonic oscillator.

\bigskip \medskip

The paper is organized as follows. Section 2 provides a brief review of the
J-matrix method and key properties of Hamiltonians with tridiagonal
representations in orthonormal bases. Section 3 presents the
tridiagonalization of the RHO, where a new matrix representation leads to
two-dimensional complex Zernike polynomials. These polynomials appear as
coefficients in the expansion of RHO-associated coherent states within the
Hilbertian probabilistic scheme. Section 4 derives a closed-form expression
for these coherent states. Section 5 verifies their Perelomov-type nature.
Finally, Section 6 applies these results by showing how the Zernike
polynomial coefficients solve the bound state problem for the Landau
Hamiltonian on the Poincar\'{e} disc. Section 7 is devoted to a summary.

\section{ The $J$-matrix method}

We assume that we are given a positive semi-definite Hamiltonian $H$ (with
zero as the value of the lowest energy in its spectrum) acting on a Hilbert
space $\mathcal{H}$, that has a tridiagonal matrix representation in the
orthonormal basis $\{{\left\vert \phi _{n}\right\rangle }\}_{n=0}^{\infty }$
of $\mathcal{H}$ with known coefficients $\{a_{n},b_{n}\}_{n=0}^{\infty }$ 
\begin{equation}
{\left\langle \phi _{n}\right\vert }H{\left\vert \phi _{m}\right\rangle }%
=b_{n-1}\,\delta _{n,m+1}+\,a_{n}\,\delta _{n,m}\,+\,b_{n}\,\delta _{n,m-1}.
\tag{2.1}
\end{equation}%
We solve the energy eigenvalue equation $H{\left\vert E\right\rangle }=E{%
\left\vert E\right\rangle }$ by expanding the eigenvector ${\left\vert
E\right\rangle }$ in the basis ${\left\vert \phi _{n}\right\rangle }$ as ${%
\left\vert E\right\rangle }=\sum_{n=0}^{\infty }f_{n}(E){\left\vert \phi
_{n}\right\rangle }$. Making use of the tridiagonality of $H$, we readily
obtain the following recurrence relations for the expansion coefficients:%
\begin{equation*}
{Ef_{0}(E)=a_{0}f_{0}(E)+b_{0}f_{1}(E)\quad }
\end{equation*}%
\begin{equation*}
{Ef_{n}(E)=b_{n-1}f_{n-1}(E)+a_{n}f_{n}(E)+b_{n}f_{n+1}(E),\quad n=1,2,...,.}
\end{equation*}%
The operator $H$ may admit a continuous spectrum $\sigma _{c}$ and a
discrete part $\{E_{\mu }\}_{\mu }$, both of which lead to the following
form of the resolution of the identity operator on $\mathcal{H}$: 
\begin{equation}
\sum_{\mu }{\left\vert E_{\mu }\right\rangle }{\left\langle E_{\mu
}\right\vert }+\int_{\sigma _{c}}{\left\vert E\right\rangle }{\left\langle
E\right\vert }\,dE\;=1_{\mathcal{H}}.\quad \quad  \tag{2.3}
\end{equation}%
This translates into the following orthogonality relation for the
coefficients $\{f_{n}\}$ : 
\begin{equation}
\sum_{\mu }f_{n}(E_{\mu })\left( f_{m}(E_{\mu })\right) ^{\ast
}+\int_{\sigma _{c}}f_{n}(E)\left( f_{m}(E)\right) ^{\ast }dE\,=\delta
_{n,m}.  \tag{2.4}
\end{equation}%
If we now define $p_{n}(E)=\frac{f_{n}(E)}{f_{0}(E)}$, then $\{p_{n}(E)\}$
is a set of polynomials that satisfy the three-term recursion relation 
\begin{equation}
Ep_{n}(E)=b_{n-1}p_{n-1}(E)+a_{n}p_{n}(E)+b_{n}p_{n+1}(E),\quad n=1,2,..., 
\tag{2.5}
\end{equation}%
with the initial conditions $p_{0}(E)=1$ and $p_{1}(E)=(E-a_{0})b_{0}^{-1}$.
If we further define $\Omega (E)=\left\vert f_{0}(E)\right\vert ^{2}$ and $%
\Omega _{\mu }=\left\vert f_{0}(E_{\mu })\right\vert ^{2}$, then the
relation $\left( 2.4\right) $ now translates into the following
orthogonality relation for the polynomial $p_{n}$: 
\begin{equation}
\sum_{\mu }\Omega _{\mu }p_{n}(E_{\mu })\left( p_{m}(E_{\mu })\right) ^{\ast
}+\int\limits_{\sigma _{c}}\Omega (E)p_{n}(E)\left( p_{m}\left( E\right)
\right) ^{\ast }dE\,=\delta _{n,m}.  \tag{2.6}
\end{equation}

We have shown $\left[ 8\right] $ that we can write the Hamiltonian $H$ in
the form $H=A^{\dag }A$, where the forward-shift operator $A$ is defined by
its action on the basis vector as 
\begin{equation}
A{\left\vert \phi _{n}\right\rangle }=c_{n}\,{\left\vert \phi
_{n}\right\rangle }+\,d_{n}\,{\left\vert \phi _{n-1}\right\rangle }. 
\tag{2.7}
\end{equation}%
Furthermore, we require from the adjoint operator $A^{\dag }$ to act on the
ket vectors ${\left\vert \phi _{n}\right\rangle }$ in the following way 
\begin{equation}
A^{\dagger }{\left\vert \phi _{n}\right\rangle }=c_{n}^{\ast }\,{\left\vert
\phi _{n}\right\rangle }+\,d_{n+1}^{\ast }\,{\left\vert \phi
_{n+1}\right\rangle .}  \tag{2.8}
\end{equation}%
Here, the coefficients $\{c_{n},d_{n}\}_{n=0}^{\infty }$ are related to the
coefficients $\{a_{n},b_{n}\}_{n=0}^{\infty }$ and the polynomials $%
\{p_{n}\}_{n=0}^{\infty }$ 
\begin{equation}
{d_{0}=0,\,\;c_{n}}c_{n}^{\ast }{=-b_{n}\frac{p_{n+1}(0)}{p_{n}(0)},\quad
d_{n+1}d_{n+1}^{\ast }=-b_{n}\frac{p_{n}(0)}{p_{n+1}(0)},}\text{ }{a_{n}=c}%
_{n}c_{n}^{\ast }+{d_{n}d_{n}^{\ast },\quad b_{n}=c_{n}d_{n+1}^{\ast }}. 
\tag{2.9}
\end{equation}

\section{Tridiagonal forms for the radial harmonic oscillator}

In this section, we first start from a known tridiagonal form of the RHO
within a basis $\phi _{n}^{\left( \lambda \right) }$ depending on a free
parameter $\lambda .$ Next, we allow $\lambda $ to take complex values and
we consider it as a function of a new parameter $\mathfrak{z}$ according to
the relation $\left( 3.6\right) $ below. This re-parametrization provides us
with a new expression for coefficients $\left( a_{n},\text{ }b_{n}\right) $
defined by $\left( 2.5\right) $ in the previous general setting. Therefore,
by the $J$-method within these coefficients

\subsection{A known tridiagonal form for the RHO}

The Hamiltonian of the radial harmonic oscillator (RHO) is

\begin{equation}
H_{\ell ,\omega }=-\frac{1}{2}\frac{d^{2}}{dr^{2}}+\frac{\ell \left( \ell
+1\right) }{2r^{2}}+\frac{1}{2}\omega ^{2}r^{2}-\left( \ell +\frac{3}{2}%
\right) \omega ,\text{ \ \ }r\in \left[ 0,+\infty \right) ,  \tag{3.1}
\end{equation}%
$\omega >0$ being a fixed parameter and $\ell =0,1,2,...,$is the angular
quantum momentum number. The space $C_{0}^{\infty }\left( \mathbb{R}%
_{+}\right) $ of infinitely differentiable complex valued functions on $%
\mathbb{R}_{+}$ with compact support is initially assumed to be the domain
of definition of $H_{\ell ,\omega }$. This space lies dense in $L^{2}\left( 
\mathbb{R}_{+}\right) $ and $H_{\ell ,\omega }$ is formally adjoint to
itself. To make $H_{\ell ,\omega }$ self-adjoint, one may appeal to its
Friedrichs extension $\cite{RS}$ which \ exists and will be denoted by $%
H_{\ell ,\omega }.$ For this extension, the spectrum in the Hilbert space $%
L^{2}\left( \mathbb{R}_{+}\right) $ is the disjoint union of its essential
spectrum $\sigma _{e}\left( H_{\ell ,\omega }\right) $ and the discrete
spectrum $\sigma _{d}\left( H_{\ell ,\omega }\right) $ which consists on a
set of eigenvalues $\varepsilon _{s}=2\omega s,$ $s=0,1,2,...,$ with the
corresponding eigenfunctions 
\begin{equation}
f_{s}\left( r\right) =\sqrt{\frac{2n!}{\Gamma \left( n+\ell +\frac{3}{2}%
\right) }}\sqrt{\omega }^{\ell +\frac{3}{2}}r^{\ell +1}e^{-\frac{1}{2}\omega
\xi ^{2}}L_{n}^{\left( \ell +\frac{1}{2}\right) }\left( \omega r^{2}\right) ,%
\text{ \ \ }r\in \mathbb{R}_{+},  \tag{3.2}
\end{equation}%
given in terms of the Laguerre polynomial \cite{Mag}, satisfying the
Dirichlet boundary condition $f_{s}\left( 0\right) =0$. The Hilbert space $%
L^{2}\left( \mathbb{R}_{+}\right) $ is the direct sum of eigenspaces of $%
H_{\ell ,\omega }$ if and only if $\sigma _{e}\left( H_{\ell ,\omega
}\right) $ is empty. This last condition can established by proving the
completeness for the system $\left\{ f_{s}\right\} _{s=0}^{\infty }$ in $%
L^{2}\left( \mathbb{R}_{+}\right) $ for $H_{\ell ,\omega }.$

We have shown in \cite{YM1} that the above Hamiltonian has a tridiagonal
matrix representation in the following orthonormal basis%
\begin{equation}
\phi _{n}^{\left( \lambda \right) }\left( r\right) =\sqrt{\frac{2\lambda n!}{%
\Gamma \left( n+\ell +\frac{3}{2}\right) }}\left( \lambda r\right) ^{\ell
+1}e^{-\frac{1}{2}\lambda ^{2}r^{2}}L_{n}^{\left( \ell +\frac{1}{2}\right)
}\left( \lambda ^{2}r^{2}\right) ,  \tag{3.3}
\end{equation}%
$\lambda $ being a free parameter. Explicitly,%
\begin{equation}
H_{\ell ,\omega }\phi _{n}^{\left( \lambda \right) }=b_{n-1}\phi
_{n-1}^{\left( \lambda \right) }+a_{n}\phi _{n}^{\left( \lambda \right)
}+b_{n}\phi _{n+1}^{\left( \lambda \right) },  \tag{3.4}
\end{equation}%
where%
\begin{equation}
a_{n}=c_{n}^{2}+d_{n}^{2},\text{ }b_{n}=c_{n}d_{n+1},\text{ \ }c_{n}=\left(
\lambda -\frac{\omega }{\lambda }\right) \sqrt{\frac{n+\ell +\frac{3}{2}}{2}}%
,\text{ }d_{n}=\left( \lambda +\frac{\omega }{\lambda }\right) \sqrt{\frac{n%
}{2}}.  \tag{3.5}
\end{equation}

\subsection{A new tridiagonal form for RHO}

Since the basis in $\left( 3.3\right) $ is labeled by the free parameter $%
\lambda ,$ we may re-labeled it by a new complex parameter $\mathfrak{z}$
such that%
\begin{equation}
\mathfrak{z:}=\sqrt{\frac{\ell +\frac{3}{2}}{2}}\left( \lambda -\frac{\omega 
}{\lambda }\right) .  \tag{3.6}
\end{equation}%
Setting $\beta =\ell +\frac{3}{2}$, the coefficients \textbf{\ }$\left\{
a_{n},b_{n}\right\} $ in $\left( 3.5\right) $ are of the form

\begin{equation}
a_{n}\equiv a_{n}^{\mathfrak{z,\omega ,\ell }}=2n\left( \omega +\frac{%
\left\vert \mathfrak{z}\right\vert ^{2}}{\beta }\right) +\left\vert 
\mathfrak{z}\right\vert ^{2}  \tag{3.7}
\end{equation}

\begin{equation}
b_{n}\equiv b_{n}^{\mathfrak{z,\omega ,\ell }}=c_{n}d_{n+1}^{\ast }=z\sqrt{%
n\left( 2\omega +\frac{\left\vert \mathfrak{z}\right\vert ^{2}}{\beta }%
\right) \left( \frac{n}{\beta }+1\right) }.  \tag{3.8}
\end{equation}%
Next, we proceed by fixing an eigenfunction $f_{m}\left( r\right) $ in $%
\left( 3.2\right) $ and expanding it over elements of the basis $\left\{
\phi _{n}^{\left( \mathfrak{z}\right) }\right\} _{n\geq 0}$ as%
\begin{equation}
f_{m}(r)=\sum_{n=0}^{\infty }\left[ \varrho \left( \varepsilon _{m}\right)
P_{n}\left( \varepsilon _{m}\right) \right] \phi _{n}^{\left( \mathfrak{z}%
\right) }\left( r\right) =\sum_{n=0}^{\infty }\left[ \gamma _{n,m}\right]
\phi _{n}^{\left( \mathfrak{z}\right) }\left( r\right) ,\text{ \ }r\in 
\mathbb{R}_{+}  \tag{3.9}
\end{equation}%
with $P_{n}$ being an orthogonal polynomial satisfying the conditions 
\begin{equation}
\sum_{m=0}^{\infty }\varrho \left( \varepsilon _{m}\right) P_{n}\left(
\varepsilon _{m}\right) \left( \varrho \left( \varepsilon _{m}\right)
P_{j}\left( \varepsilon _{m}\right) \right) ^{\ast }=\delta _{n,j},\text{ \ }%
P_{0}\left( \varepsilon _{m}\right) =1.  \tag{3.10}
\end{equation}%
where $\varrho $ stands for the square root of an orthonormalization density
function. From the tridiagonality condition $\left( 3.4\right) $, we know
that $p_{n}$ satisfy the following three-term recurrence relation

\begin{equation}
\varepsilon _{m}P_{n}\left( \varepsilon _{s}\right) =b_{n-1}P_{n-1}\left(
\varepsilon _{m}\right) +a_{n}P_{n}\left( \varepsilon _{m}\right)
+b_{n}^{\ast }P_{n+1}\left( \varepsilon _{m}\right) .  \tag{3.11}
\end{equation}%
By $\left( 3.9\right) $-$\left( 3.10\right) $, Eq.$\left( 3.11\right) $
takes the form 
\begin{equation}
\varepsilon _{m}P_{n}\left( \varepsilon _{m}\right) =\mathfrak{z}\sqrt{%
\left( n-1\right) \left( 2\omega +\frac{\left\vert \mathfrak{z}\right\vert
^{2}}{\beta }\right) \left( \frac{n-1}{\beta }+1\right) }P_{n-1}\left(
\varepsilon _{m}\right) +\left( 2n\left( \omega +\frac{\left\vert \mathfrak{z%
}\right\vert ^{2}}{\beta }\right) +\left\vert \mathfrak{z}\right\vert
^{2}\right) P_{n}\left( \varepsilon _{s}\right)  \tag{3.12}
\end{equation}%
\begin{equation*}
+\mathfrak{z}^{\ast }\sqrt{n\left( 2\omega +\frac{\left\vert \mathfrak{z}%
\right\vert ^{2}}{\beta }\right) \left( \frac{n}{\beta }+1\right) }%
P_{n+1}\left( \varepsilon _{s}\right) .
\end{equation*}%
We introduce the following change of polynomial by setting 
\begin{equation}
P_{n}\left( u\right) =\left( -1\right) ^{n}\sqrt{\frac{\left( \beta \right)
_{n}}{n!}}\left( \left\vert \mathfrak{z}\right\vert ^{2}+2\beta \omega
\right) ^{-\frac{1}{2}n}\left( \mathfrak{z}^{\ast }\right) ^{n}\mathfrak{Q}%
_{n}\left( u;\left\vert \mathfrak{z}\right\vert ^{2}\right) ,\text{ \ }%
u\equiv \varepsilon _{m}  \tag{3.13}
\end{equation}%
where $u\mapsto \mathfrak{Q}_{n}\left( u;\left\vert \mathfrak{z}\right\vert
^{2}\right) $ is a polynomial, depending on the parameter $\left\vert 
\mathfrak{z}\right\vert ^{2}=\mathfrak{zz}^{\ast }$, to be determined. For
this we insert $\left( 4.5\right) $ into $\left( 4.4\right) ,$ to obtain the
following equation

\begin{equation*}
\varepsilon _{m}\mathfrak{Q}_{n}\left( \varepsilon _{m};\left\vert \mathfrak{%
z}\right\vert ^{2}\right) =-\left\vert \mathfrak{z}\right\vert ^{2}\left( 
\frac{n}{\beta }+1\right) \mathfrak{Q}_{n+1}\left( \varepsilon ;\left\vert 
\mathfrak{z}\right\vert ^{2}\right) +\left( \left\vert \mathfrak{z}%
\right\vert ^{2}+n\left( 2\omega +\frac{2\left\vert \mathfrak{z}\right\vert
^{2}}{\beta }\right) \right) \mathfrak{Q}_{n}\left( \varepsilon
_{m};\left\vert \mathfrak{z}\right\vert ^{2}\right)
\end{equation*}%
\begin{equation}
-n\left( 2\omega +\frac{\left\vert \mathfrak{z}\right\vert ^{2}}{\beta }%
\right) \mathfrak{Q}_{n-1}\left( \varepsilon _{m};\left\vert \mathfrak{z}%
\right\vert ^{2}\right) .\text{\ }  \tag{3.14}
\end{equation}%
From the latter one we recognize the three-term recurrence relation of the
Meixner polynomials ($\cite{KLS}):$

\begin{equation}
\mathfrak{Q}_{n}\left( \varepsilon _{m};\left\vert z\right\vert ^{2}\right)
=M_{n}\left( m,\beta ;\frac{\left\vert z\right\vert ^{2}}{\left\vert
z\right\vert ^{2}+2\beta \omega }\right)  \tag{3.15}
\end{equation}%
which are defined by the general formula 
\begin{equation}
M_{n}\left( x,\tau ;c\right) =\left( -1\right)
^{n}\sum\limits_{k=0}^{n}\left( 
\begin{array}{c}
x \\ 
k%
\end{array}%
\right) \left( 
\begin{array}{c}
-x-\tau \\ 
n-k%
\end{array}%
\right) c^{-k},\text{ }n=0,1,2,...\text{ .}  \tag{3.16}
\end{equation}%
Consequently,%
\begin{equation}
P_{n}\left( \varepsilon _{m}\right) =\left( -1\right) ^{n}\sqrt{\frac{\left(
\beta \right) _{n}}{n!}}\left( \left\vert \mathfrak{z}\right\vert
^{2}+2\beta \omega \right) ^{-\frac{1}{2}n}\left( \mathfrak{z}^{\ast
}\right) ^{n}M_{n}\left( m,\beta ;\frac{\left\vert \mathfrak{z}\right\vert
^{2}}{\left\vert \mathfrak{z}\right\vert ^{2}+2\beta \omega }\right) 
\tag{3.17}
\end{equation}%
being an orthogonal polynomial satisfying the conditions 
\begin{equation}
\sum_{m=0}^{\infty }\left( \frac{\left( \beta \right) _{m}}{m!}\left( \frac{%
\mathfrak{z}}{\left\vert \mathfrak{z}\right\vert ^{2}+2\beta \omega }\right)
^{m-\beta }\right) ^{\frac{1}{2}}P_{n}\left( \varepsilon _{m}\right) \left(
\left( \frac{\left( \beta \right) _{m}}{m!}\left( \frac{\mathfrak{z}}{%
\left\vert \mathfrak{z}\right\vert ^{2}+2\beta \omega }\right) ^{m-\beta
}\right) ^{\frac{1}{2}}P_{j}\left( \varepsilon _{m}\right) \right) ^{\ast
}=\delta _{n,j},\text{ \ }P_{0}\left( \varepsilon _{j}\right) =1.  \tag{3.18}
\end{equation}%
Finally, back to the expansion $\left( 3.9\right) ,$ the coefficients we
were seeking for are of the form%
\begin{equation*}
\gamma _{n,m}=\left( \frac{\left( \beta \right) _{m}}{m!}\left( \frac{%
\mathfrak{z}}{\left\vert \mathfrak{z}\right\vert ^{2}+2\beta \omega }\right)
^{m-\beta }\right) ^{\frac{1}{2}}\left( -1\right) ^{n}\sqrt{\frac{\left(
\beta \right) _{n}}{n!}}\left( \left\vert \mathfrak{z}\right\vert
^{2}+2\beta \omega \right) ^{-\frac{1}{2}n}\left( \mathfrak{z}^{\ast
}\right) ^{n}M_{n}\left( m,\beta ;\frac{\left\vert \mathfrak{z}\right\vert
^{2}}{\left\vert \mathfrak{z}\right\vert ^{2}+2\beta \omega }\right)
\end{equation*}%
\begin{equation}
=\left( \frac{\left\vert \mathfrak{z}\right\vert ^{2}}{\left\vert \mathfrak{z%
}\right\vert ^{2}+2\beta \omega }\right) ^{-\frac{1}{2}\beta }C_{n,m}\left( 
\mathfrak{z},\mathfrak{z}^{\ast }\right)  \tag{3.19}
\end{equation}

\subsection{Link with Zernike polynomials}

In order to link the above obtained polynomials $C_{n,m}\left( \mathfrak{z},%
\mathfrak{z}^{\ast }\right) $ in $\left( 3.19\right) $ \ with the well known
disc polynomials, we need to rewrite the following expression%
\begin{equation}
C_{n,m}\left( \mathfrak{z},\mathfrak{z}^{\ast }\right) =\left( -1\right) ^{n}%
\sqrt{\frac{\left( \beta \right) _{m}\left( \beta \right) _{n}}{m!n!}}\left(
\left\vert \mathfrak{z}\right\vert ^{2}+2\beta \omega \right) ^{-\frac{1}{2}%
m-\frac{1}{2}n}\mathfrak{z}^{\ast n}\mathfrak{z}^{m}M_{n}\left( m,\beta ;%
\frac{\left\vert \mathfrak{z}\right\vert ^{2}}{\left\vert \mathfrak{z}%
\right\vert ^{2}+2\beta \omega }\right)   \tag{3.20}
\end{equation}%
in a more recognizable form. For that, we first recall that the Meixner
polynomials may also be expressed in terms of Jacobi polynomials as $\left( 
\cite{KLS}\right) :$%
\begin{equation}
M_{n}\left( x,\tau ;c\right) =\left( -c\right) ^{-n}P_{n}^{\left( x-n,\tau
-1\right) }\left( 1-2c\right) ,\text{ }n=0,1,2,...,  \tag{3.21}
\end{equation}%
Therefore, $\left( 3.20\right) $ transforms, after some calculations, to

\begin{equation}
C_{n,s}\left( \mathfrak{z},\mathfrak{z}^{\ast }\right) =\left( -1\right) ^{n}%
\sqrt{\frac{\left( \beta \right) _{s}\left( \beta \right) _{n}}{s!n!}}\left( 
\frac{\mathfrak{z}}{\sqrt{\left\vert \mathfrak{z}\right\vert ^{2}+2\beta
\omega }}\right) ^{s-n}P_{n}^{\left( \beta -1,s-n\right) }\left( 2\frac{%
\mathfrak{zz}^{\ast }}{\sqrt{\left\vert z\right\vert ^{2}+2\beta \omega }%
\sqrt{\left\vert z\right\vert ^{2}+2\beta \omega }}-1\right)   \tag{3.22}
\end{equation}%
This last expression, naturally, suggest us to introduce the new complex
variable $z$ in the unit disc $\mathbb{D}$, defined by%
\begin{equation}
z:=\frac{\mathfrak{z}}{\sqrt{\left\vert \mathfrak{z}\right\vert ^{2}+2\beta
\omega }}.  \tag{3.23}
\end{equation}%
Therefore, 
\begin{equation}
C_{n,s}\left( \mathfrak{z},\mathfrak{z}^{\ast }\right) =\left( -1\right) ^{n}%
\sqrt{\frac{\left( \beta \right) _{s}\left( \beta \right) _{n}}{s!n!}}%
z^{s-n}P_{n}^{\left( \beta -1,s-n\right) }\left( 2zz^{\ast }-1\right) . 
\tag{3.24}
\end{equation}%
Note that, we have here recovered the well known generalized Zernike of $%
\left( z,z^{\ast }\right) $ of degree $m+n$ 
\begin{equation}
P_{s,n}^{\beta -1}\left( z,z^{\ast }\right) :=\frac{n!\Gamma \left( \beta
\right) }{\Gamma \left( n+\beta \right) }z^{s-n}P_{n}^{\left( \beta
-1,s-n\right) }\left( 2zz^{\ast }-1\right) ==\frac{s!\Gamma \left( \beta
\right) }{\Gamma \left( s+\beta \right) }z^{\ast n-s}P_{n}^{\left( \beta
-1,n-s\right) }\left( 2zz^{\ast }-1\right)   \tag{3.25}
\end{equation}%
which also admit a second basic explicit representation 
\begin{equation}
P_{s,n}^{\alpha }\left( z,z^{\ast }\right) =\sum\limits_{k=0}^{s\wedge n}%
\frac{\left( -1\right) ^{k}s!n!\alpha !}{k!\left( s-k\right) !\left(
n-k\right) !\left( k+\alpha \right) !}\left( 1-zz^{\ast }\right)
^{k}z^{s-k}z^{\ast n-k}.  \tag{3.26}
\end{equation}%
Properties of these polynomials can be found in \cite{Wun} and references
therein.

\section{A closed form for $\protect\phi _{n}^{\left( \mathfrak{z}\right)
}\equiv \protect\phi _{n}^{\left( z\right) }$}

Using the orthogonality relation $\left( 3.18\right) $, we can invert the
relation $\left( 3.9\right) $ in order to expand a fixed vector $\phi
_{n}^{\left( \mathfrak{z}\right) }$ over the eigenvectors basis $\left(
f_{s}\right) _{s\geq 0}$ as follows%
\begin{equation}
\phi _{n}^{\left( \mathfrak{z}\right) }=\sum\limits_{s=0}^{\infty }\left[
\eta _{s,n}\left( \mathfrak{z}\right) \right] ^{\ast }f_{s}  \tag{4.1}
\end{equation}%
where%
\begin{equation}
\eta _{s,n}\left( \mathfrak{z}\right) =\left( -1\right) ^{n}\sqrt{\frac{%
n!\left( \beta \right) _{s}}{s!\left( \beta \right) _{n}}}\left( \sqrt{\tau }%
\right) ^{s-n}\left( 1-\xi \right) ^{\beta /2}P_{n}^{\left( \beta
-1,s-n\right) }\left( 2\xi -1\right)   \tag{4.2}
\end{equation}%
with%
\begin{equation}
\sqrt{\tau }=\frac{\mathfrak{z}}{\sqrt{\left\vert \mathfrak{z}\right\vert
^{2}+2\beta \omega }},\text{ \ \ }\beta =\ell +\frac{3}{2},\text{ \ }\xi
=\left\vert \tau \right\vert =\frac{\left\vert \mathfrak{z}\right\vert ^{2}}{%
\left\vert \mathfrak{z}\right\vert ^{2}+2\beta \omega }  \tag{4.3}
\end{equation}%
Setting 
\begin{equation}
z=\frac{\mathfrak{z}}{\sqrt{\left\vert \mathfrak{z}\right\vert ^{2}+2\beta
\omega }}\in \mathbb{D}  \tag{4.4}
\end{equation}%
Then $\left( \sqrt{\tau }\right) ^{\ast }=z^{\ast },\xi =zz^{\ast
}=\left\vert z\right\vert ^{2}.$With these notations, Eq. $(4.2)$ becomes%
\begin{equation}
\left[ \eta _{s,n}\left( \mathfrak{z}\right) \right] ^{\ast }=\left(
-1\right) ^{n}\sqrt{\frac{n!\left( \beta \right) _{s}}{s!\left( \beta
\right) _{n}}}\left( z^{\ast }\right) ^{s-n}\left( 1-\left\vert z\right\vert
^{2}\right) ^{\beta /2}P_{n}^{\left( \beta -1,s-n\right) }\left( 2\left\vert
z\right\vert ^{2}-1\right)   \tag{4.5}
\end{equation}%
Recalling the expression $\left( 3.2\right) $ of eigenfunctions $f_{s},$ we
seek for a closed for the series $\left( 4.1\right) $:%
\begin{equation}
S=\frac{\left( -1\right) ^{n}\left( \sqrt{\omega }\right) ^{\beta }}{\sqrt{%
\Gamma \left( \beta \right) }}\sqrt{\frac{2n!}{\left( \beta \right) _{n}}}%
\left( z^{\ast }\right) ^{-n}\left( 1-\left\vert z\right\vert ^{2}\right)
^{\beta /2}r^{\ell +1}e^{-\frac{1}{2}\omega r^{2}}  \tag{4.6}
\end{equation}%
\begin{equation*}
\times \sum\limits_{s=0}^{\infty }\left( z^{\ast }\right) ^{s}P_{n}^{\left(
\beta -1,s-n\right) }\left( 2\left\vert z\right\vert ^{2}-1\right)
L_{s}^{\left( \beta -1\right) }\left( \omega r^{2}\right) 
\end{equation*}%
Make use of the connection formula $\left( \cite{Bry}\text{, p.556, Eq.17}%
\right) $ :%
\begin{equation}
P_{n}^{\left( \alpha ,s-n\right) }\left( X\right) =\frac{s!}{n!}\frac{\Gamma
\left( n+\alpha +1\right) }{\Gamma \left( s+\alpha +1\right) }\left( \frac{%
X+1}{2}\right) ^{n-s}P_{s}^{\left( \alpha ,n-s\right) }\left( X\right)  
\tag{4.7}
\end{equation}%
the last sum in $\left( 4.6\right) $ takes the form%
\begin{equation}
\frac{zz^{\ast }}{n!}\left( \beta \right) _{n}\sum\limits_{s=0}^{\infty
}\left( \frac{1}{z}\right) ^{s}\frac{s!}{\left( \beta \right) _{s}}%
P_{s}^{\left( \beta -1,n-s\right) }\left( 2\left\vert z\right\vert
^{2}-1\right) L_{s}^{\left( \beta -1\right) }\left( \omega r^{2}\right) . 
\tag{4.8}
\end{equation}%
By making use of the generating function $\left( \cite{Prud2}\text{, p.718}%
\right) :$%
\begin{equation}
\sum\limits_{s=0}^{\infty }u^{s}\frac{s!}{\left( \alpha +1\right) _{s}}%
P_{s}^{\left( \alpha ,n-s\right) }\left( Y\right) L_{s}^{\left( \alpha
\right) }\left( X\right)   \tag{4.9}
\end{equation}%
\begin{equation*}
=\left( 1-u\right) ^{n}\left( 1-\left( \frac{1+Y}{2}\right) u\right)
^{-\alpha -n-1}\exp \left( \frac{u}{u-1}X\right) ._{1}\digamma _{1}\left( 
\begin{array}{c}
\alpha +n+1 \\ 
\alpha +1%
\end{array}%
;\frac{uX\left( 1-Y\right) }{\left( 1-u\right) \left( 2-u-uY\right) }\right)
,
\end{equation*}%
we obtain, after calculations, the wave function of our basis $\phi
_{n}^{\left( \mathfrak{z}\right) }\equiv \phi _{n}^{\left( z\right) }$ 
\begin{equation}
\langle r\left\vert \phi _{n,\beta ,\omega }^{\left( z\right) }\right\rangle
=\left( \sqrt{\omega }r\right) ^{\ell +1}\left( \frac{2\sqrt{\omega }\left(
\beta \right) _{n}}{n!\Gamma \left( \beta \right) }\right) ^{\frac{1}{2}%
}\left( \frac{1-z}{1-z^{\ast }}\right) ^{n}\left( 1-z^{\ast }\right)
^{2-\beta }\left( 1-zz^{\ast }\right) ^{\beta /2}\exp \left( \frac{1}{2}%
\omega r^{2}\left( \frac{1+z}{1-z}\right) \right)   \tag{4.10}
\end{equation}%
\begin{equation*}
\times _{1}\digamma _{1}\left( \beta +n,\beta ;-\frac{\left( 1-zz^{\ast
}\right) }{\left\vert 1-z\right\vert ^{2}}\omega r^{2}\right) 
\end{equation*}%
Next, using the identity $\left( \cite{Prud3}\text{, p. 579}\right) :$%
\begin{equation}
_{1}\digamma _{1}\left( a,a-n;X\right) =\frac{\left( -1\right) ^{n}n!}{%
\left( 1-a\right) _{n}}e^{X}L_{n}^{a-n-1}\left( -X\right)   \tag{4.11}
\end{equation}%
we may write the $_{1}F_{1}$ as%
\begin{equation}
\frac{\left( -1\right) ^{n}n!}{\left( 1-\beta -n\right) _{n}}\exp \left( -%
\frac{\left( 1-zz^{\ast }\right) }{\left\vert 1-z\right\vert ^{2}}\omega
r^{2}\right) L_{n}^{\beta -1}\left( \frac{\left( 1-zz^{\ast }\right) }{%
\left\vert 1-z\right\vert ^{2}}\omega r^{2}\right) .  \tag{4.12}
\end{equation}%
Therefore Eq. $\left( 4.10\right) $ reads

\begin{equation*}
\langle r\left\vert \phi _{n,\beta ,\omega }^{\left( z\right) }\right\rangle
=\left( \sqrt{\omega }r\right) ^{\ell +1}\frac{\left( -1\right) ^{n}n!}{%
\left( 1-\beta -n\right) _{n}}\left( \frac{2\sqrt{\omega }\left( \beta
\right) _{n}}{n!\Gamma \left( \beta \right) }\right) ^{\frac{1}{2}}\left( 
\frac{1-z}{1-z^{\ast }}\right) ^{n}\left( 1-zz^{\ast }\right) ^{\beta /2}
\end{equation*}%
\begin{equation}
\times \exp \left( -\frac{\left( 1-zz^{\ast }\right) }{\left\vert
1-z\right\vert ^{2}}\omega r^{2}+\frac{1}{2}\omega r^{2}\left( \frac{1+z}{1-z%
}\right) \right) L_{n}^{\beta -1}\left( \frac{\left( 1-zz^{\ast }\right) }{%
\left\vert 1-z\right\vert ^{2}}\omega r^{2}\right)  \tag{4.13}
\end{equation}%
Finally,%
\begin{equation*}
\langle r\left\vert \phi _{n,\beta ,\omega }^{\left( z\right) }\right\rangle
=\left( \sqrt{\omega }r\right) ^{\ell +1}\frac{\left( -1\right) ^{n}\sqrt{n!}%
}{\left( 1-\beta -n\right) _{n}}\left( \frac{2\sqrt{\omega }\left( \beta
\right) _{n}}{\Gamma \left( \beta \right) }\right) ^{\frac{1}{2}}\left( 
\frac{1-z}{1-z^{\ast }}\right) ^{n}\left( 1-zz^{\ast }\right) ^{\beta
/2}\left( 1-z^{\ast }\right) ^{-\beta }
\end{equation*}

\begin{equation}
\times \exp \left( -\frac{1}{2}\omega r^{2}\left( \frac{1+z^{\ast }}{%
1-z^{\ast }}\right) \right) L_{n}^{\beta -1}\left( \frac{\left( 1-zz^{\ast
}\right) }{\left\vert 1-z\right\vert ^{2}}\omega r^{2}\right) .  \tag{4.14}
\end{equation}%
Note that in the case $n=0,$the first element of this basis should be
coherent state. \ Indeed, since $\beta =\ell +\frac{3}{2},$ we have the
following wavefunction%
\begin{equation}
\langle r\left\vert \phi _{0,\beta ,\omega }^{\left( z\right) }\right\rangle
=\left( \frac{2\sqrt{\omega }}{\Gamma \left( \ell +\frac{3}{2}\right) }%
\right) ^{\frac{1}{2}}\left( \sqrt{\omega }r\right) ^{\ell +1}\frac{\left(
1-\left\vert z\right\vert ^{2}\right) ^{\frac{1}{2}\ell +\frac{3}{4}}}{%
\left( 1-z^{\ast }\right) ^{\ell +\frac{3}{2}}}\exp \left( -\frac{1}{2}%
\omega r^{2}\left( \frac{1+z^{\ast }}{1-z^{\ast }}\right) \right) ,\text{ \
\ \ \ }z\in \mathbb{D}  \tag{4.15}
\end{equation}%
which, exactly, is the wave function of the radial coherent states $\left( 
\cite{GK},\text{ p.193, Eq. (3.13)}\right) $.

\medskip

For $n\geq 1,$ we will prove that $\phi _{n,\beta ,\omega }^{\left( z\right)
}$ are generalized coherent states of Perelomov-type, which may also be
considered as displaced Fock states where the displacement operator is
played by an operator of a square integrable representation of the affine
group. This will be the aim of the next section.

\section{ $\left( \protect\phi _{n}^{\left( z\right) }\right) _{z\in \mathbb{%
D}}$ as Perelomov's coherent states}

The affine group is the set $\mathbf{G}:=\mathbb{R}\times \mathbb{R}^{+}$,
endowed with group law $\left( x,y\right) \cdot \left( x^{\prime },y^{\prime
}\right) =\left( x+yx^{\prime },yy^{\prime }\right) $. $\mathbf{G}$ is a
locally compact group with the left Haar measure $d\mu \left( x,y\right)
=y^{-2}dxdy$. \ Here, we will be concerned with the following continuous
unitary irreducible (UIR) dimensional UIR, denoted $T_{\omega }$, acting on
the Hilbert space $\mathfrak{H}=$ $L^{2}\left( \mathbb{R}^{+},dr\right) ,$
and is given by $\left( \cite{Thei}\right) :$ 
\begin{equation}
T_{\omega }\left( x,y\right) \left[ \varphi \right] \left( r\right) :=y^{%
\frac{1}{4}}\exp \left( -\frac{1}{2}ix\omega r^{2}\right) \varphi \left( 
\sqrt{y}r\right)  \tag{5.1}
\end{equation}%
for $\left( x,y\right) \in \mathbf{G}$ and $\varphi \in \mathfrak{H}.$ This
representation is square integrable since it is easy to find a vector $\Phi
\in \mathfrak{H}$ such that the function $\left( x,y\right) \mapsto
\left\langle T_{\omega }\left( x,y\right) \left[ \Phi \right] ,\Phi
\right\rangle _{\mathfrak{H}}$ belongs to $L^{2}\left( \mathbf{G},d\mu
\right) $. Such a vector is called $T$-admissible. \ This condition is,
here, equivalent to 
\begin{equation}
\gamma :=2\pi \int\limits_{\mathbb{R}^{+}}\left\vert \Phi \left( r\right)
\right\vert ^{2}r^{-2}dr<+\infty .  \tag{5.2}
\end{equation}%
By choosing as reference state (or fuducial vector) the function (Fock
state) $f_{n}$ given by $\left( 3.2\right) $, this constant is given by 
\begin{equation}
\gamma :=2\pi \int\limits_{\mathbb{R}^{+}}\left\vert f_{n}\left( r\right)
\right\vert ^{2}r^{-2}dr  \tag{5.3}
\end{equation}%
\begin{equation}
=\frac{4\pi \omega ^{\ell +\frac{3}{2}}n!}{\Gamma \left( n+\ell +\frac{3}{2}%
\right) }\int\limits_{0}^{+\infty }r^{2\ell }e^{-\omega r^{2}}\left(
L_{n}^{\left( \ell +\frac{1}{2}\right) }\left( \omega r^{2}\right) \right)
^{2}dr=\frac{2\pi \omega }{\ell +\frac{1}{2}}<+\infty .  \tag{5.4}
\end{equation}%
The coherent states are defined by displacing the Fock state $f_{n}$ via the
operator $T_{\omega }\left( x,y\right) $ as 
\begin{equation}
\left\vert \tau _{(x,y),l,\omega ,n}\right\rangle :=T_{\omega }\left(
x,y\right) \left[ f_{n}\right] .  \tag{5.5}
\end{equation}%
Taking into account the coefficient $\left( 5.3\right) $, the identity
operator on $\mathfrak{H}$ decomposes as 
\begin{equation}
\mathbf{1}_{\mathfrak{H}}=\frac{1}{\gamma }\int\limits_{\mathbf{G}%
}y^{-2}dxdy\left\vert \tau _{(x,y),l,\omega ,n}\right\rangle \left\langle
\tau _{_{(x,y),l,\omega ,n}}\right\vert  \tag{5.6}
\end{equation}%
where the Dirac's bra-ket notation $|\Psi \rangle \langle \Psi |$ means the
rank-one operator $\phi \longmapsto \langle \Psi ,\phi \rangle _{\mathfrak{H}%
}.\Psi $ with $\Psi ,\phi \in \mathfrak{H}$.

\medskip \medskip

The wave functions of the coherent states $\left( 5.5\right) $ in the $\xi $%
-coordinate read%
\begin{equation}
\langle r\left\vert \tau _{(x,y),l,\omega ,n}\right\rangle =\sqrt{\frac{2n!}{%
\Gamma \left( n+\ell +\frac{3}{2}\right) }}\left( \sqrt{\omega }\right)
^{\ell +\frac{3}{2}}y^{\frac{1}{2}\ell +\frac{3}{4}}r^{1+\ell }\exp \left( -%
\frac{1}{2}\omega (ix+y)r^{2}\right) L_{n}^{\left( \ell +\frac{1}{2}\right)
}\left( y\omega r^{2}\right) ,\text{ }\left( x,y\right) \in \mathbf{G}\text{.%
}  \tag{5.7}
\end{equation}%
By using the inverse Cayley transform $\mathcal{C}^{-1}:\mathbb{D}%
\rightarrow \mathbf{G}$ given by 
\begin{equation}
\mathcal{C}^{-1}\left( z\right) =\left( \frac{-2\func{Im}z}{\left\vert
1-z\right\vert ^{2}},\frac{1-\left\vert z\right\vert ^{2}}{\left\vert
1-z\right\vert ^{2}}\right) ,\text{ \ }z\in \mathbb{D}\text{,}  \tag{5.8}
\end{equation}%
we may define a version of these states in the same Hilbert space $\mathfrak{%
H}$ as follows%
\begin{equation}
\left\vert \tau _{z,l,\omega ,n}\right\rangle :=\left( \frac{1-z}{1-z^{\ast }%
}\right) ^{\frac{1}{2}\left( \ell +\frac{3}{2}\right) }T_{\omega }\left( 
\mathcal{C}^{-1}\left( z\right) \right) \left[ f_{n}\right] .  \tag{5.9}
\end{equation}%
Their wave functions are now labeled by points of the\ $z\in \mathbb{D}$
instead of point $\left( x,y\right) \in \mathbf{G}$ as 
\begin{equation}
\langle r\left\vert \tau _{z,l,\omega ,n}\right\rangle =\left( \frac{n!2%
\sqrt{\omega }}{\Gamma \left( n+\ell +\frac{3}{2}\right) }\right) ^{\frac{1}{%
2}}\left( \sqrt{\omega }r\right) ^{\ell +1}\frac{\left( 1-\left\vert
z\right\vert ^{2}\right) ^{\frac{1}{2}\ell +\frac{3}{4}}}{\left( 1-z^{\ast
}\right) ^{\ell +\frac{3}{2}}}\exp \left( -\frac{1}{2}\omega r^{2}\left( 
\frac{1+z^{\ast }}{1-z^{\ast }}\right) \right) L_{n}^{\left( \ell +\frac{1}{2%
}\right) }\left( \omega \left( \frac{1-\left\vert z\right\vert ^{2}}{%
\left\vert 1-z\right\vert ^{2}}\right) r^{2}\right)  \tag{5.10}
\end{equation}%
which is the same as the obtained ones in $\left( 4.14\right) $.

\section{\textbf{An application to hyperbolic Landau levels}}

We start by recalling the\ GCS $\left( 5.7\right) $ in $L^{2}(\mathbb{R}%
_{+},dr)$ which were labeled by points $\left( x,y\right) $ of the affine
group $\mathbf{G.}$

\begin{equation}
\langle r\left\vert \tau _{(x,y),l,\omega ,n}\right\rangle =\sqrt{\frac{2n!}{%
\Gamma \left( n+\ell +\frac{3}{2}\right) }}\sqrt{\omega }^{\ell +\frac{3}{2}%
}y^{\frac{1}{2}\ell +\frac{3}{4}}r^{1+\ell }\exp \left( -\frac{1}{2}\omega
(ix+y)r^{2}\right) L_{n}^{\left( \ell +\frac{1}{2}\right) }\left( y\omega
r^{2}\right) ,\text{ }\left( x,y\right) \in \mathbf{G.}  \tag{6.1}
\end{equation}%
For our purpose, we would like to transfer them to the Hilbert space $L^{2}(%
\mathbb{R}_{+},r^{-1}dr).$ For that, we slightly modify them as follows 
\begin{equation}
\langle r\left\vert \tau _{(x,y),\ell ,\omega ,n}\right\rangle =\sqrt{\frac{2%
}{r}}\,\langle r^{2}|\kappa _{(x,y),\ell ,\omega ,n}\rangle .  \tag{6.2}
\end{equation}%
Indeed, $r\mapsto \langle r|\kappa _{(x,y),\ell ,\omega ,n}\rangle $ belongs
to $L^{2}(\mathbb{R}_{+},r^{-1}d\xi ),$and is given by 
\begin{equation}
\langle r|\kappa _{(x,y),\ell ,\omega ,n}\rangle =\sqrt{\frac{n!}{\Gamma
\left( n+\ell +\frac{3}{2}\right) }}\left( \sqrt{\omega }\right) ^{\ell +%
\frac{3}{2}}y^{\frac{1}{2}\ell +\frac{3}{4}}r^{\frac{1}{2}(\frac{3}{2}+\ell
)}\exp \left( -\frac{1}{2}\omega (ix+y)r\right) L_{n}^{\left( \ell +\frac{1}{%
2}\right) }\left( y\omega r\right) .  \tag{6.3}
\end{equation}%
Here $\ell ,\omega $ and $n$ are fixed\ parameters. We introduce a new
parameter $B$ such that%
\begin{equation}
2B=\ell +\frac{3}{2}+2n\   \tag{6.4}
\end{equation}%
\text{\ and we rewrite Eq. }$\left( 6.3\right) $ in terms of $B$ as

\begin{equation}
\langle r|\kappa _{(x,y),B,\omega ,n}\rangle =\sqrt{\frac{n!}{\Gamma \left(
2B-n\right) }}\omega ^{B-n}(yr)^{B-2n}\exp \left( -\frac{1}{2}\omega
(ix+y)r\right) L_{n}^{\left( 2(B-n)-1\right) }\left( y\omega r\right) . 
\tag{6.5}
\end{equation}%
To describe the connection of CS $\left( 6.5\right) $ with the $n$th
hyperbolic Landau level, we first identify the affine group $\mathbf{G\equiv 
}\mathbb{H}^{2}$ with the Poincar\'{e} upper half-plane $\mathbb{H}%
^{2}=\left\{ x+iy,\text{ }x\in \mathbb{R},y>0\right\} $. Then, to these CS
we attach, as usual, the coherent state transform $\mathcal{B}_{n}:L^{2}(%
\mathbb{R}_{+},r^{-1}dr)\rightarrow L^{2}\left( \mathbb{H}^{2},d\mu \right) $
defined by $\cite{Mou1}$: 
\begin{equation}
\mathcal{B}_{n}[\phi ]\left( x,y\right) =c_{B,n}\int\limits_{0}^{+\infty
}\left\langle r\right\vert \kappa _{(x,y),B,\omega ,n}\rangle \phi
(r)r^{-1}dr  \tag{6.6}
\end{equation}%
where $c_{B,n}:=\sqrt{2\left( B-n\right) -1}$. The range of $\mathcal{B}_{n}$
coincides with the eigenspace%
\begin{equation}
\mathcal{E}_{B,n}:=\left\{ \digamma \in L^{2}\left( \mathbb{H}^{2},d\mu
\right) ,\;H_{B}\digamma =\epsilon _{n}^{B}\digamma \right\}  \tag{6.7}
\end{equation}%
of the Schr\"{o}dinger operator\ (in suitable units and up to an additive
constant) describing the dynamics of a charged particle moving on $\mathbb{H}%
^{2}$ under the action of a magnetic field of strength proportional to $B:$ 
\begin{equation}
H_{B}=y^{2}\left( \partial _{x}^{2}+\partial _{y}^{2}\right) -2iBy\partial
_{x},  \tag{6.8}
\end{equation}%
associated with the eigenvalue 
\begin{equation}
\epsilon _{n}^{B}=(B-n)\left( 1-B+n\right) ,\quad n=0,1,...,\left\lfloor B-%
{\frac12}%
\right\rfloor ,  \tag{6.9}
\end{equation}%
where $\left\lfloor a\right\rfloor $ denotes the greatest integer not
exceeding $a.$ We precisely have $\mathcal{B}_{n}[L^{2}(\mathbb{R}%
_{+},r^{-1}dr)]=\mathcal{E}_{B,n}.$ The operator $H_{B}$ an elliptic densely
defined operator on the Hilbert space $L^{2}(\mathbb{H}^{2},d\mu )$, with a
unique self-adjoint realization also denoted by $H_{B}$. Its spectrum
consists of two parts:\textit{\ }a continuous part $\left[ 
{\frac14}%
,+\infty \right[ $, corresponding to scattering states and the finite number
of eigenvalues $\epsilon _{n}^{B}$ each one with infinite degeneracy,\
called hyperbolic Landau levels \cite{Com}.

\medskip

Next, we transform the Schr\"{o}dinger operator $H_{B}$ into a form that
acts on functions defined on the complex unit disc $\mathbb{D}$. To achieve
this, we employ the Cayley transform $\mathcal{C}:$ $\mathbb{H}%
^{2}\rightarrow \mathbb{D}$, $w\mapsto \left( w-i\right) \left( w+i\right)
^{-1}.$ Precisely, if $g:\mathbb{H}^{2}\rightarrow \mathbb{C}$ is an
eigenfunction of $\left( -H_{B}\right) $ with $\lambda $ as eigenvalue, then
the function 
\begin{equation}
\mathbb{D}\ni z\mapsto \left( \frac{\overline{\mathcal{C}^{-1}\left(
z\right) }-i}{\mathcal{C}^{-1}\left( z\right) +i}\right) ^{B}g\left(
C^{-1}\left( z\right) \right) ,\text{ \ }  \tag{6.10}
\end{equation}%
is an eigenfunction of the Schr\"{o}dinger operator with magnetic field on $%
\mathbb{D}$, given by

\begin{equation}
\Delta _{B}=\frac{-1}{4}\left( \left( 1-zz^{\ast }\right) ^{2}\frac{\partial
^{2}}{\partial z\partial z^{\ast }}+Bz\left( 1-zz^{\ast }\right) \frac{%
\partial }{\partial z}-Bz^{\ast }\left( 1-zz^{\ast }\right) \frac{\partial }{%
\partial z^{\ast }}+B^{2}\left( 1-zz^{\ast }\right) \right)   \tag{6.11}
\end{equation}%
Therefore, the eigenspace $\mathcal{E}_{B,n}$ in $\left( 6.7\right) $
translates into the following 
\begin{equation}
\mathcal{A}_{B,n}=\left\{ \Phi :\mathbb{D\rightarrow C},\text{ }\Phi \in
L^{2}\left( \mathbb{D},\left( 1-\left\vert z\right\vert ^{2}\right) ^{-2}%
\frac{d\nu \left( z\right) }{\pi }\right) ,\text{ }\Delta _{B}\Phi =\epsilon
_{n}^{B}\Phi \right\} ,  \tag{6.12}
\end{equation}%
where $d\nu \left( z\right) $ is the Lebesgue measure on $\mathbb{D}$. An
orthogonal basis of this eigenspace is given the functions $\left( \cite%
{Mou2}\right) $:%
\begin{equation}
\Phi _{j}^{B,n}\left( z\right) :=\left\vert z\right\vert ^{\left\vert
j\right\vert }\left( 1-zz^{\ast }\right) ^{B-n}e^{-ij\theta
}._{2}F_{1}\left( -n+\frac{1}{2}\left( j+\left\vert j\right\vert \right)
,2B-n+\frac{1}{2}\left( \left\vert j\right\vert -j\right) ,1+\left\vert
j\right\vert ;\left\vert z\right\vert ^{2}\right) ,\text{ }-\infty \leq 
\text{\ }j\leq n\text{,}  \tag{6.13}
\end{equation}%
with the norms square

\begin{equation}
\left\Vert \Phi _{j}^{B,n}\right\Vert ^{2}=\frac{\left( \left\vert
j\right\vert !\right) ^{2}}{\left( 2\left( B-n\right) -1\right) }\frac{%
\Gamma \left( n-\frac{1}{2}\left( \left\vert j\right\vert +j\right)
+1\right) }{\Gamma \left( n+\frac{1}{2}\left( \left\vert j\right\vert
-j\right) +1\right) }\frac{\Gamma \left( 2B-n-\frac{1}{2}\left( \left\vert
j\right\vert +j\right) \right) }{\Gamma \left( 2B-n+\frac{1}{2}\left(
\left\vert j\right\vert -j\right) \right) }.  \tag{6.14}
\end{equation}%
By Setting $n-j=s\geq 0,$ we have two cases:%
\begin{equation}
\Phi _{n-s}^{B,n}\left( z\right) =\left\{ 
\begin{array}{c}
\left( 1-\left\vert z\right\vert ^{2}\right) ^{B-n}\frac{s!}{n!}\left(
n-s\right) !\left( z^{\ast }\right) ^{n-s}P_{s}^{\left( n-s,2(B-n)-1\right)
}\left( 1-2\left\vert z\right\vert ^{2}\right) ,\text{ }n\geq s \\ 
\left( 1-\left\vert z\right\vert ^{2}\right) ^{B-n}\frac{n!}{s!}\left(
s-n\right) !z^{s-n}P_{n}^{\left( s-n,2(B-n)-1\right) }\left( 1-2\left\vert
z\right\vert ^{2}\right) ,\text{ }n\leq s%
\end{array}%
\right.  \tag{6.15}
\end{equation}%
Recalling that $2B=\beta +2n$, we may choose case $s\geq n$ \ in order to
re-write $\left( 6.16\right) $ in terms of $\beta $ as \ 
\begin{equation}
\Phi _{n-s}^{B,n}\left( z\right) =\left( 1-\left\vert z\right\vert
^{2}\right) ^{\frac{1}{2}\beta }\frac{n!}{s!}\left( s-n\right)
!z^{s-n}P_{n}^{\left( s-n,\beta -1\right) }\left( 1-2\left\vert z\right\vert
^{2}\right) .  \tag{6.16}
\end{equation}%
Next, by using the symmetry relation $\left( \cite{Bry}\text{, p.555}\right)
:$ 
\begin{equation}
P_{m}^{\left( a,b\right) }\left( -X\right) =\left( -1\right)
^{m}P_{m}^{\left( b,a\right) }\left( X\right) ,  \tag{6.17}
\end{equation}%
we obtain the normalized basis vector of $\ $the eigenspace $\mathcal{A}%
_{B,n}$ as 
\begin{equation}
\frac{\Phi _{n-s}^{B,n}\left( z\right) }{\left\Vert \Phi
_{n-s}^{B,n}\right\Vert }=c_{B,n}\left( 1-\left\vert z\right\vert
^{2}\right) ^{^{\frac{1}{2}\beta }}\left( -1\right) ^{n}\sqrt{\frac{n!\left(
\beta \right) _{s}}{s!\left( \beta \right) _{n}}}z^{s-n}P_{n}^{\left( \beta
-1,s-n\right) }\left( 2\left\vert z\right\vert ^{2}-1\right)  \tag{6.18}
\end{equation}%
where $c_{B,n}==\sqrt{2\left( B-n\right) -1}$ already appeared as a constant
in $\left( 6.6\right) .$Finally, by recalling the expansion $\left(
4.1\right) $ 
\begin{equation*}
\phi _{n}^{\left( \mathfrak{z}\right) }=\sum\limits_{s=0}^{\infty }\left[
\eta _{s,n}\left( \mathfrak{z}\right) \right] ^{\ast }f_{s}
\end{equation*}%
where 
\begin{equation*}
\eta _{s,n}\left( \mathfrak{z}\right) \equiv \eta _{s,n}\left( z\right)
=\left( -1\right) ^{n}\sqrt{\frac{n!\left( \beta \right) _{s}}{s!\left(
\beta \right) _{n}}}z^{s-n}\left( 1-\left\vert z\right\vert ^{2}\right)
^{\beta /2}P_{n}^{\left( \beta -1,s-n\right) }\left( 2\left\vert
z\right\vert ^{2}-1\right)
\end{equation*}%
we finally arrive at the precise result 
\begin{equation}
\frac{\Phi _{n-s}^{B,n}\left( z\right) }{\left\Vert \Phi
_{n-s}^{B,n}\right\Vert }=c_{B,n}\left[ \eta _{s,n}\left( z\right) \right] ,%
\text{ \ \ \ }z\in \mathbb{D}\text{.}  \tag{6.19}
\end{equation}%
This confirms that the bound state solutions of the two-dimensional Schr\"{o}%
dinger operator $\Delta _{B}$ are given by the expansion coefficients of the
basis elements $\phi _{n}^{\left( \mathfrak{z}\right) }$ (which
tridiagonalize the radial harmonic oscillator (RHO) and depend on the unit
disc parameter $z$) when expressed in the eigenstate basis $\left(
f_{s}\right) $ of the RHO.

\bigskip

\section{Summary}

This work explores the tridiagonalization of the radial harmonic oscillator
(RHO) Hamiltonian using the $J$-matrix method, constructing its matrix
representation in an orthonormal basis $\phi _{n}^{\left( z\right) }$ of $%
L^{2}\left( \mathbb{R}_{+}\right) ,$ parametrized by a fixed $z$ in\ the
complex unit disc $\mathbb{D}$ and $n=0,1,2,...$ . For a fixed $n$, we
demonstrate that the expansion coefficients $c_{n,s}(z,\overline{z})$ of $%
\phi _{n}^{\left( z\right) }$ over RHO eigenstates basis $\left\{
f_{s}\right\} $ coincide with two-dimensional complex Zernike polynomials.
Crucially, we establish that $\phi _{n}^{\left( z\right) }$ (for varying $z$
in $\mathbb{D}$) forms a set of Perelomov-type coherent states, obtained by
the displacement action of operators from a square-integrable representation
of the affine group on an RHO eigenstate (Fock state). This justifies their
interpretation as displaced Fock states. A central result is the encoding of
bound states for the two-dimensional Schr\"{o}dinger operator on the Poincar%
\'{e} disc within the RHO's tridiagonal structure: superpositions of
coefficients $c_{n,s}(z,\overline{z})$\ generate eigenstates of hyperbolic
Landau levels with magnetic strength $B>%
{\frac12}%
$ , offering a hyperbolic counterpart to prior Euclidean solutions via the
cross-Wigner transform. \ While our previous work \cite{YM2} established the 
$J$-matrix method for Euclidean geometry (harmonic oscillator/complex
Hermite polynomials), this paper demonstrates its extension to hyperbolic
geometry through the radial harmonic oscillator and Zernike polynomials on
the Poincar\'{e} disc. The natural next step (applying this approach to
spherical geometry via Kravchuk oscillators) would solve the Dirac monopole
problem on the Riemann sphere, thereby completing the geometric trilogy and
proving the $J$-matrix method's universality for quantum systems with
magnetic fields across all fundamental geometries.

\end{document}